\documentstyle[11pt]{article}
\def\p{\partial}   
  
\def\e{\epsilon} \def\g{\gamma} \def\n{\nabla} \def\d{\delta}
\def\r{\rho}    \def\eb{{\bar \eta}}
\def\x{\chi}   \def\bt{{\bar \vartheta}}
\def\b{\beta} \def\a{\alpha} \def\l{\lambda}  \def\f{\varphi}
\def\da{{\dot \alpha}} \def\db{{\dot \beta}}  \def\dg{{\dot \gamma}}
\def\dd{{\dot \delta}} \def\de{{\dot \eta}}
 
\def\sd{self-dual }
\def\sdym{self-dual Yang-Mills }

\def\ssd{super self-dual }
\def\sdy{self-duality } \def\ssdy{super self-duality }
\def\sym{super Yang-Mills } \def\ym{Yang-Mills }
\def\reps{representations }
\def\susy{supersymmetric }\def\eqs{equations }
  
  \def\eom{equation of motion }
\def\eoms{equations of motion }
\def\half{{1\over 2}} \def\N#1{$N=#1$ } \def\Tr{\mbox {Tr }}
\def\der#1{{\partial \over \partial #1}}

\def\be{\begin{equation}}
\def\r#1{(\ref{#1})} \def\la#1{\label{#1}} \def\c#1{\cite{#1}}
\def\ee{\end{equation}}
\def\arr{\begin{array}{rll}}
\def\ea{\end{array}}
\def\bea{\begin{eqnarray}}
\def\eea{\end{eqnarray}}
\def\square{\kern1pt\vbox
            {\hrule height 0.6pt\hbox{\vrule width 0.6pt\hskip 3pt
  \vbox{\vskip 6pt}\hskip 3pt\vrule width 0.6pt}\hrule height 0.6pt}\kern1pt}

\begin{document}
\rightline{hep-th/9510235}
\vskip 1 true cm
\centerline{       CONSERVED CURRENTS FOR UNCONVENTIONAL        }
\centerline{ SUPERSYMMETRIC COUPLINGS OF SELF-DUAL GAUGE FIELDS }
\vskip 1 true cm
\centerline{        Ch. Devchand  and  V. Ogievetsky }
\vskip .35 true cm
\centerline{{\it     Joint Institute for Nuclear Research,
141980 Dubna, Russia}}
\vskip 1 true cm
\noindent
{\bf Abstract: }
Self-dual gauge potentials admit supersymmetric couplings to higher-spin
fields satisfying interacting forms of the first order Dirac--Fierz equation.
The interactions are governed by conserved currents determined by
supersymmetry. These super-self-dual Yang-Mills systems provide on-shell
supermultiplets of arbitrarily extended super-Poincar\'e algebras; classical
consistency not setting any limit on the extension N. We explicitly display
equations of motion up to the $N=6$ extension. The stress tensor, which
vanishes for the $N\le 3$ self-duality equations, not only gets resurrected
when $N=4$, but is then a member of a conserved multiplet of gauge-invariant
tensors.
\vskip 20pt
{\bf 1. }
In the standard maximally supersymmetric \N4  \ym theory \c{bss}, both the
\sd $(1,0)$ as well as the anti-\sd $(0,1)$ parts of the \ym field strength
are contained in the {\it same} supermultiplet. This is not the case in lower
$N$ theories, where these two halves of the field strength live in separate
irreducible representations of the supersymmetry algebra which, although
conjugate in Minkowski space, are independent in spaces having other
signatures. This means that pure \sym theories with $N\le 3$ admit \ssd
restriction, i.e. systems of equations which include the \ym \sdy condition
$f_{\da\db}=0$, which are invariant under the N-extended super Poincar\'e
algebra, and which imply the full set of \sym equations. The standard \N4
theory does not admit such a \ssd restriction. However there does exist a
rather remarkable \N4 \susy extension of the \sdy condition \c{s}, which was
inspired by string theory \c{ov}. This \sd theory is
independent of the standard maximally supersymmetric \ym theory \c{bss}.
Not only are the  \eoms of the latter theory not implied, but the
spectrum of fields differs. The \N4 \sd theory contains an
additional spin 1 field, independent of the \ym vector potential. It is
remarkable that gauge invariance allows such a coupling to the vector
potential. In standard \ym theory,
$$ \e^{\da\dg} \p_{\a\dg} f_{\da\db} + \e^{\b\g} \p_{\g\db} f_{\a\b}
    =\ J_{\a\db}\   ,$$
the conserved vector current $J_{\a\db}$ provides {\it all} consistent
spin 1 couplings \c{op}.
In the absence of any gauge invariances beyond the \ym one, massless
\sym multiplets contain, with the exception of the gauge potential
$A_{\a\db}$, only fields transforming according to the either the $(s,0)$
or the $(0,s)$ \reps of the rotation group, and according to skew-symmetric
representations of the internal $SL(N)$ automorphism group of the
N-extended supersymmetry algebra.
The \ssdy equations up to \N3 take the following N-independent forms
\be\arr
 f_{\da\db}& =& 0 \cr
 \e^{\a\g} \n_{\g\db} \l_{i\a}& =\  & 0 \cr
 \n^{\a\db}\n_{\a\db} W_{ij} &  =\  & \{ \l^\a_i, \l_{j\a} \} \cr
 \e^{\da\db} \n_{\g\db} \x_{ijk\da }& =\  &  [ \l_{[i\a} , W_{jk]} ]
  ,\la{c3}\ea\ee
where $\n_{\g\db} =  \p_{\g\db} + A_{\g\db}$ is the gauge--covariant
derivative and we have scaled the gauge coupling constant to one;
all fields are gauge algebra valued and are therefore linear in the
coupling constant. They are skew-symmetric in the internal $sl(N)$ indices
$i,j = 1,...,N $, which we always write as subscripts \footnote{All our
(skew-)symmetrisations are with weight one. For instance,
$ [ \l_{[i\a} , W_{jk]} ] \equiv [ \l_{i\a} , W_{jk}] +
[ \l_{j\a} , W_{ki}] + [ \l_{k\a} , W_{ij}]  $}. This is not always the most
economical description of the degrees of freedom, for instance, there is only
one scalar for \N2,  $W_{ij}\equiv \e_{ij} W$, or for \N3, three scalars,
$W_{ij}\equiv \e_{ijk} W^k $, and one $(0,\half)$ spinor,
$ \x_{ijk\da}\equiv  \e_{ijk}\x_\da$. However, this notation, which we use
throughout this paper, has the advantage of being N-independent.

The equations \r{c3} imply the full \eoms of the standard \sym theories
\c{bss}. They also display an interesting nested structure \c{matr}; the
fields $W_{ij}$ which exist for $N \ge 2$ do not occur in the $N < 2$ \sdy
\eqs, and the fields $\x_{ijk\da }$ which appear when \N3 do not occur in
the lower N \eoms. This nested structure, which led us previously to call
this system a {\it \sd matreoshka}, is crucial for our present discussion.
For in fact the {\it matreoshka} is even larger because of the following.

\noindent {\it If we allow the internal indices to range over four values,
$i,j,k= 1,\dots ,4$, then equations \r{c3} imply that the following
vector current is covariantly conserved: }
\be  J_{ijkl \a\db}  =  \{ \l_{[i\a} , \x_{jkl]\db} \}
                       -  [ W_{[ij}, \n_{\a\db} W_{kl]} ]\   ,\la{v}\ee
$$ \mbox{i.e.}\quad  \n^{\a\db} J_{ijkl \a\db} = 0\  .$$
This current affords the enhancement of the \N3
multiplet to an \N4 one by the addition of a spin 1 field $g_{ijkl\da\db}$
satisfying the \eom
\be \e^{\da\db} \n_{\a\db} g_{ijkl\da\dg } =\  J_{ijkl \a\dg}\    ;\la{c4}\ee
all the previous \eoms \r{c3} remaining intact. This is precisely the \N4
\sd theory presented by Siegel. The Lorentz--covariant functional
\be S_{ijkl} = \int d^4x\  \Tr \left( f^{\da\db}g_{ijkl\da\db}
               + \x_{[ijk}^\da \n_{\a\da} \l^\a_{l]}
	       + \half W_{[ij}\n^{\a\db}\n_{\a\db}W_{kl]}
	       -  W_{[ij} \{ \l^\a_{k} , \l_{l]\a} \}
	       \right)  ,\la{s4}\ee
is an $sl(4)$ singlet and provides an action $S = \e^{ijkl}S_{ijkl}$
for the \N4 theory \c{s}.
The two conserved vector currents, from the \eoms of the two spin 1 fields
$A_{\a\db}$ and $g_{\da\db}$, are manifestly independent:
\be\arr
 j^{(1)}_{\a\db} =\  &  - \p^\da_\a [A^\g_{(\da}, A_{\g\db)}]\    ,\cr
 j^{(2)}_{ijkl \a\db} =\ & ( J_{ijkl \a\db}
                   - \e^{\da\dg} [ A_{\a\dg}, g_{ijkl\da\db}] )\  .\ea\ee
Conservation of the latter, i.e. that
$$ \p^{\a\db} j^{(2)}_{ijkl\a\db} = 0\   ,$$
corresponds to the global gauge--invariance of the functional \r{s4},
whereas conservation of the former can be interpreted as a
consequence of the global gauge--invariance of the pure \ym
functional when the \sdy conditions are satisfied.

The field equations (\ref{c3},\ref{c4}), with internal indices now taken
to range over five values, similarly imply the covariant constancy of an
\N5 current
\be  J_{ijklm \a\da\db}  =  [ \l_{[i\a} , g_{jklm]\da\db} ]
        + {2\over 3}[\n_{\a(\da} W_{[ij},  \x_{klm]\db)} ]
	- {1\over 3}[ W_{[ij}, \n_{\a(\da} \x_{klm]\db)} ]\    ,\la{v5}\ee
affording the enhancement of the system (\ref{c3},\ref{c4}) by the spin
${3\over 2}$ equation
\be \e^{\da\db}\n_{\a\db}\psi_{ijklm\da\dg\dd } =\
                                    J_{ijklm \a\dg\dd}\  .\la{c5}\ee
The gauge--covariant conservation of $J_{ijklm \a\dg\dd}$ implies the
existence of a non--gauge--covariant divergence--free spin--vector current,
$$
j_{ijklm \a\dg\dd} =\ J_{ijklm \a\dg\dd} -
		      \e^{\da\db}[A_{\a\db}, \psi_{ijklm\da\dg\dd}]\  .$$
In turn, an  \N6 spin 2 field can be introduced, with \eom
\be\arr \e^{\da\db}\n_{\a\db}C_{ijklmn\da\dg\dd\de}
& = & \{ \l_{[i\a} , \psi_{jklmn]\dg\dd\de} \}
	+ {1\over 6}\{ \x_{[ijk(\dg}, \n_{\a\dd} \x_{lmn]\de)} \}
  \\[4pt]
       & & + \half [\n_{\a(\dg} W_{[ij},  g_{klmn]\dd\de)} ]
        - {1\over 6}[ W_{[ij}, \n_{\a(\dg} g_{klmn]\dd\de)} ]
         .\la{c6}\ea\ee
Again, the right-hand-side is a covariantly conserved current. This pattern
continues; and it actually continues {\it ad infinitum}, essentially because
the $(N-1)$--extended system nestles within the $N$--extended system
completely intact, and provides a conserved source current for a new spin
${(N-2)\over 2}$ field of dimension $-{N\over 2}$. This is a further
unconventional feature of these \sd theories: The dimensions of our fields
depend linearly on the spin, whereas in conventional field theories all
bosons have dimension $-1$ and all fermions have dimension $-{3\over 2}$.
It is tempting to speculate on the significance of the infinite number of
local conserved currents in the infinite $N$ limit of a theory with
infinitely many spins reminicent of string theories.

The higher spin--vector conserved currents
$J_{i_1\dots i_N \a\da_1 \dots \da_{(N-2)}} $ may be obtained from the vector
currents \r{v} by performing supersymmetry transformations. They therefore
also essentially owe their existence to the gauge--invariant functionals
$S_{ijkl}$, which are extremised by solutions of the \eoms even for $N>4$,
in spite of the fact that higher spin fields manifestly do not contribute to
them. The supersymmetry transformations of the \N6 equations are given by
\be\begin{array}{lll}  \d A_{\a\db}&=& - \eb^i_\db  \l_{i\a}    \\[4pt]
   \d \l_{j\a} & = &  \eta_j^\b  f_{\a\b}
                       + 2 \eb^{i\db}  \n_{\a\db} W_{ij}      \\[4pt]
   \d W_{jk}   & = &  \eta_{[j}^\a \l_{k]\a}
                       + \eb^{i\db} \x_{ijk\db}     \\[4pt]
 \d \x_{jkl\da} & = &  \eta_{[j}^\a \n_{\a\da} W_{kl]}
    + \eb^{i\db} (g_{ijkl\da\db} + \e_{\da\db} [W_{i[j},W_{kl]} ] ) \\[4pt]
 \d g_{jklm\da\db} & = &  \eta_{[j}^\a \n_{\a(\da} \x_{klm]\db)}    \\[4pt]
  && + \eb^{i\dg}(\psi_{ijklm\da\db\dg}
  +\e_{\dg(\da} ( {2\over 3}[W_{i[j},\x_{klm]\db)}]
                  - {1\over 3}[W_{[jk},\x_{lm]i\db)}]) )
     \\[4pt]
 \d \psi_{jklmn\da\db\dg} & = &  \eta_{[j}^\a \n_{\a(\da} g_{klmn]\db\dg)}
               + \eb^{i\dd} C_{ijklmn\da\db\dg\dd} \\[4pt]
     && - \eb^i_{(\da} ( \half [W_{i[j}, g_{klmn]\db\dg)}]
                      + {1\over 6}[W_{[jk}, g_{lmn]i\db\dg)}]
	      + {1\over 6} [\x_{i[jk\db}, \x_{lmn]\dg)}]  )    \\[4pt]
 \d C_{jklmnp\da\db\dg\dd} &=&
\eta_{[j}^\a \n_{\a(\da} \psi_{klmnp]\db\dg)}\\[4pt]
        && - \eb^i_{(\da} ( {2\over 5} [W_{i[j}, \psi_{klmnp]\db\dg\dd)}]
                   - {1\over 10} [W_{[jk}, \psi_{lmnp]i\db\dg\dd)}] \\[4pt]
       &&\qquad  - {1\over 15} [ \x_{[jkl\db}, g_{mnp]i\dg\dd)} ]
- {1\over 10} [ \x_{i[jk\db}, g_{lmnp]\dg\dd)} ] )\     .\la{susy}\ea\ee
These are most conveniently obtained from the superspace formulation of
these theories, where superfield versions of the functionals $S_{ijkl}$
exist, generalising the \N4 superspace actions \c{s,e}. Our arbitrarily
extended \sdy equations may be compactly expressed in terms of the chiral
superspace curvature constraints
\be\arr
\{\n_{i \da}, \n_{j \db}\} =& \e_{\da \db} f_{i j} \\[2pt]
[\n_{i \da}, \n_{\b \db}] =& \e_{\da \db}   f_{i \b}   \\[2pt]
[\n_{\a \da}, \n_{\b \db}] =& \e_{\da \db}  f_{\a \b}\  ,
\la{constr}\ea\ee where
$ \n_{i \da} = \der{\bt_{i \da}} + A_{i \da}  ,\quad
  \n_{\a \da} = \der{x_{\a \da}} + A_{\a\da} $ are chiral superspace
gauge-covariant derivatives. We shall present a proof of the equivalence
of these constraints to the equations of motion (\ref{c3}, \ref{c4},
\ref{c5}, \ref{c6}) in a separate publication.
\vskip 10pt
{\bf 2. }
For \N0 \sdy the stress tensor vanishes identically
$$ T_{\alpha{\dot \alpha},\beta{\dot \beta}}
\equiv\ \Tr f_{{\dot \alpha}{\dot \beta}} f_{\alpha\beta} \equiv\ 0 .$$
This remains true for all supersymmetrisations up to $N=3$. However,
the appearance of the invariant functional \r{s4} for \N4 resurrects the
stress tensor. For $N\ge4$, in fact, there exist $\pmatrix{ N\cr 4 }$
second rank traceless (i.e. satisfying
$\e^{\a\b}\e^{\da\db} T_{ijkl\a\da ,\b\db} =0 $) conserved tensors,
corresponding to this number of invariant functionals $S_{ijkl}$ :
\be \arr
T_{ijkl\a\da ,\b\db} &=& \Tr (
     g_{ijkl\da\db} f_{\a\b}  +  \n_{\a\db} \l_{[i\b} \x_{jkl]\da}
   -    \l_{[i\a} \n_{\b\da} \x_{jkl]\db}                \\[4pt]
 && \qquad  + {1\over 2} \l_{[i\b} \n_{\a\da} \x_{jkl]\db}
        - {1\over 2} \n_{\a\da} \l_{[i\b} \x_{jkl]\db}
 + {2\over 3} \e_{\db\da}\e_{\b\a} \{ \l^\g_{[i}, \l_{j\g}\} W_{kl]} \\[4pt]
 && \qquad   - {1\over 3} \n_{(\a\da} W_{[ij}\n_{\b)\db} W_{kl]}
             + {1\over 3} W_{[ij} \n_{\a\da}\n_{\b\db} W_{kl]} )  .\la{tr}
    \ea\ee
In fact there exist three second rank conserved tensors,
\be\arr
T^{(1)}_{ijkl\a\da ,\b\db}&=& \Tr (
   g_{ijkl\da\db} f_{\a\b} -  \l_{[i\a} \n_{\b\da} \x_{jkl]\db}
   -  \n_{\b\da} W_{[ij}\n_{\a\db} W_{kl]}  ) \\[4pt]
T^{(2)}_{ijkl\a\da ,\b\db} &=& \Tr (
 {1\over 2} \l_{[i\b} \n_{\a\da} \x_{jkl]\db}
 - {1\over 2} \n_{\a\da} \l_{[i\b} \x_{jkl]\db}
 +  \n_{\a\db} \l_{[i\b} \x_{jkl]\da}  \\[4pt] &&\qquad
 +  \e_{\b\a} \e_{\db\da} \{ \l^\g_{[i}, \l_{j\g} \} W_{kl]}  ) \\[4pt]
T^{(3)}_{ijkl\a\da ,\b\db} &=& \Tr {1\over 3}(
    W_{[ij} \n_{\a\da}\n_{\b\db} W_{kl]}
    -  \n_{\a\da} W_{[ij} \n_{\b\db} W_{kl]}
    + 2 \n_{\b\da} W_{[ij} \n_{\a\db} W_{kl]}   \\[4pt]  &&\qquad
    - \e_{\b\a} \e_{\db\da} \{ \l^\g_{[i}, \l_{j\g} \} W_{kl]}  )\   ,\ea\ee
of which the sum \r{tr} is the unique traceless linear combination.
(The $\Tr$ in these expressions denotes of course the gauge algebra trace).
These gauge--invariant tensors have conserved superpartners. The lower rank
conserved spin--tensors are
\be\begin{array}{lll}
T_{ijk\a\da ,\b} &=& \Tr  (2 f_{\a\b} \x_{ijk\da}
   - \n_{\a\da} \l_{[i\b} W_{jk]}
   +  \l_{[i\b} \n_{\a\da} W_{jk]}
   - 2 \l_{[i\a} \n_{\b\da} W_{jk]} )   \\[4pt]
T^{(1)}_{ijklm\a\da ,\db} &=& \Tr (
     4 \l_{i\a} g_{jklm\da\db} - \l_{[j} g_{klm]i\da\db}
     - 4 \x_{i[jk\da} \n_{\a\db} W_{lm]}           \\[4pt]   &&\qquad
     + 6 \n_{\a\db} W_{i[j} \x_{klm]\da}
   - 5 \e_{\da\db} W_{i[j} [ \l_{k\a} ,W_{lm]} ]   ) \\[4pt]
T^{(2)}_{ijklm\a\da ,\db} &=& \Tr (
 \n_{\a\da}W_{i[j}\x_{klm]\db} -  W_{i[j}\n_{\a\da}\x_{klm]\db}
	    - 2 \n_{\a\db} W_{i[j}\x_{klm]\da} \\[4pt]   &&\qquad
   + 2 \e_{\da\db} W_{i[j} [ \l_{k\a} ,W_{lm]} ]   ) \\[4pt]
T^{(3)}_{ijklm\a\da ,\db} &=& \Tr (
 \n_{\a\da}\x_{i[jk\db} W_{lm]} -  \x_{i[jk\db} \n_{\a\da} W_{lm]}
	 + 2 \x_{i[jk\da} \n_{\a\db} W_{lm]} \\[4pt]   &&\qquad
   + 2 \e_{\da\db} W_{i[j} [ \l_{k\a} ,W_{lm]} ]   ) \\[4pt]
T_{ijkl\a\da} &=& \Tr  ( 3 \l_{i\a} \x_{jkl\da} + \l_{[j\a} \x_{kl]i\da}
          +  2 \n_{\a\da} W_{i[j} W_{kl]} - 2 W_{i[j}\n_{\a\da} W_{kl]}  )\
	                      .\ea\ee
All these tensors satisfy the conservation law
$$ \p^{\a\da} T_{i\dots m \a\da , \dots} =\  0   $$
in virtue of the equations of motion, and they may be used to couple these
\sd gauge theories to gravity and supergravity.
\vskip 10pt
{\bf 3. }
The free (but massive) versions of the higher spin equations
(\ref{c4}, \ref{c5}, \ref{c6}) were considered a long time ago by Dirac
\c{dirac} and by Fierz \c{fierz}; and the problems of consistently coupling
such fields to an external electromagnetic field were discussed. In the
zero rest-mass limit, corresponding to the zero coupling limit of our
equations,
\be  \p_\g^{\da_1} f_{\da_1 \dots \da_{2s}}
  =\ 0\  ,\la{free}\ee
there is no problem in consistently coupling an external {\it \sd} \ym
field to such spinors (this of course requires the background space to
have signature $(4,0)$ or $(2,2)$). The replacement of $\p_{\a\da}$
in \r{free} by the gauge--covariant derivative requires the satisfaction
of precisely the \sdy equation $f_{\da\db}=0$ for consistency.
Such a coupling of a zero rest--mass spinor to a {\it \sd} vector potential
appears to be the
unique consistent coupling of the type which Dirac attempted to find.
If a non-zero source current $J$ is present, the further consistency
condition for a minimal gauge coupling is the gauge--covariant constancy
of the current. As we have seen supersymmetric \sdym theory
automatically provides such conserved currents.
In fact all our higher spin equations (\ref{c4}, \ref{c5}, \ref{c6})
have the general form
\be  \p_\a^{\da_1} \f_{i_1\dots i_{(2s+2)} \da_1\da_2 \dots \da_{2s}}
  =\  j_{i_1\dots i_{(2s+2)} \a \da_2 \dots \da_{2s} },\quad  s\ge 1,
  \la{cur}\ee
where the current on the right is a functional of all fields of spin $\le s$
(including the \ym vector potential). Now differentiating on the left,
$$\p^{\a\da_2} \p_\a^{\da_1} \f_{i_1\dots i_{(2s+2)} \da_1\da_2\dots\da_{2s}}
= \half \e^{\da_1\da_2} \p^{\g\db}\p_{\g\db}
\f_{i_1\dots i_{(2s+2)} \da_1\da_2 \dots \da_{2s}}\  ,$$
clearly shows that the consistency condition for the linear equation \r{cur}
is precisely the conservation of the current on the right, since $\f$ is
symmetric in its dotted spinor indices whereas $\e$ is skew--symmetric.
This is analogous to the consistency requirement in conventional theories.
For instance, for zero mass vector fields, consistency of the equation
$\p^\mu F_{[\mu\nu]} = J_\nu$ implies $\p^\nu J_\nu =0$ in virtue of
the antisymmetry of $F_{\mu\nu}$.

The zero rest-mass Dirac-Fierz equation \r{free}  has also been studied
by Penrose \c{p} who discussed the possible geometrical significance of the
spin 2 case. It remains an intruiguing open question whether any relation
exists between our \N6 theory and his considerations. We note, however, that
our \N5 theory is probably the unique supersymmetric theory in which a
spin ${3\over 2}$ field is coupled to a vector field, without requiring
a spin 2 coupling as well for consistency \c{os}. Traditional theorems
forbidding higher--spin couplings do not apply to our systems since these
\sd theories have only one coupling constant (the \ym one) and only one type
of gauge--invariance (also the \ym one). Even for \N5 or \N6 there is no
further coupling constant and no additional gauge--invariance. The fields
$\psi_{\da\db\dg}$ and $C_{\da\db\dg\dd}$, being gauge algebra valued,
transform covariantly under \ym gauge transformations and have dimensions
$-{5\over 2}$ and $-3$ respectively. It is these high negative
dimensionalities which render it impossible to write down action functionals
for these higher--spin fields. We should note, however, that {\it locally}
supersymmetric versions of our higher spin theories are also possible,
having spin $1$, spin ${3\over 2}$, and spin $2$ gauge--invariances,
as well as both Yang-Mills and gravitational coupling constants. We are
currently investigating such $N \ge 8$ \sd supergravities.
\goodbreak
\end{document}